\documentclass[aps,preprint,amssymb]{revtex4}

\usepackage{epsfig}
\usepackage{lscape,graphicx}
\usepackage{rotating}
\usepackage{color}
\usepackage{amsmath}

\begin{document}
\renewcommand{\thefootnote}{\fnsymbol{footnote}} 
\renewcommand{\theequation}{\arabic{section}.\arabic{equation}}

\title{Kinesin's Backsteps under Mechanical Load}

\author{Changbong Hyeon,$^a$ Stefan Klumpp,$^b$ and Jos{\'e} N. Onuchic$^b$}
\email[]{E-mail: jonuchic@ucsd.edu}

\affiliation{$^a$ Department of Chemistry, Chung-Ang University, Seoul 156-756, Republic of Korea\\
$^b$ Center for Theoretical Biological Physics and Department of Physics, University of California at San Diego, La Jolla, California 92093-0374}

\date{\today}

\begin{abstract}
\noindent 
Kinesins move processively toward the plus end of microtubules by hydrolyzing ATP for each step. From an enzymatic perspective, the mechanism of mechanical motion coupled to the nucleotide chemistry is often well explained using a single-loop cyclic reaction. 
However, several difficulties arise in interpreting kinesin's backstepping within this framework, especially when external forces oppose the motion of kinesin.  
We review evidence, such as an ATP-independent stall force and a slower cycle time for backsteps, that has emerged to challenge the idea that kinesin backstepping is due to ATP synthesis, i.e., the reverse cycle of kinesin's forward-stepping chemomechanics. 
Supplementing the conventional single-loop chemomechanics with routes for ATP-hydrolyzing backward steps and nucleotide-free steps, especially under load, gives a better physical interpretation of the experimental data on backsteps.  
\end{abstract}

\maketitle

\section{Introduction}

Structural flexibility, ligand-protein interactions, conformational changes and free energy transduction are among the key properties that distinguish molecular nanomachines from macroscopic motors. 
Kinesins are one type of such molecular motors that beautifully coordinate these properties to unidirectionally move 
along microtubule (MT) filaments in a thermally noisy environment \cite{Vale85Cell}.
The quest to understand how those properties are integrated into a molecule of a few nm size to provide its full biological function has been pursued for  
the last two decades using careful experiments both at the ensemble \cite{Gilbert94Biochem,GilbertPNAS04,Moyer98BC,Hackney05PNAS,Hackeny94PNAS,Ma97JBC} and single molecule level \cite{Howard89Nature,Hunt94BJ,Cross07Science,Block03Science,Block03PNAS,Block90Nature,Cross05Nature,Ishiwata01Science,Coppin96PNAS,Yildiz08Cell,Yildiz04Science,Mori07Nature,ValeNature99,Gelles88Nature} along with theories \cite{Kolomeisky07ARPC,Fisher01PNAS,Fisher05PNAS,Liepelt07PRL,Hyeon07PNAS,Hyeon07PNAS2,Reimann02PR,Tsygankov07PNAS,Lipowsky08JSP,Chemla08JPCB}.
Yet, multifaceted responses to external stress, the signature of complexity and versatility of molecular architecture, often make it difficult to decipher the physics of kinesin dynamics using a single experimental probe \cite{Block07BJ}.

As one of such multifaceted responses of kinesin motors, 
recent experimental reports on the backward stepping of kinesin has drawn much attention \cite{Block07BJ,Cross05Nature,Nishiyama02NCB,Yildiz08Cell}.
Kinesins show occasional backward steps even in the absence of an opposing force, but their frequency increases exponentially under an external load opposing the movement the motor.  
Carter and Cross have demonstrated that kinesin can step backward even in a processive fashion if it experiences a load that exceeds a stall force \cite{Cross05Nature}.
It is easy to imagine physiological situations where the ability to step backward is advantageous, such as to avoid obstacles on MTs \cite{Ross08COCB}, or to assist the coordination of multiple kinesins \cite{Beeg08BJ} or kinesins and dyneins \cite{Gross04PhysBiol,Muller08PNAS} on the same cargo.
However, the mechanistic interpretation of backsteps is not straightforward: Is the backstepping cycle a reversal of the conventional forward cycle, i.e., does kinesin {\it synthesize ATP} while it steps backward under super-stall loads \cite{Block07BJ}?

Despite a plethora of different models,
the current understanding of the chemomechanics of kinesins is largely based on a single cyclic chemical reaction that simplifies the dynamics of kinesin into a series of state-to-state transitions \cite{Cross04TBS,Schief01COCB,Kolomeisky07ARPC}. 
In this article, we review the experimental data on the backsteps of kinesin and point out the puzzles associated with interpreting these data within the conventional picture of {\it a single} chemomechanical cycle. We argue that these puzzles are resolved if the backward steps occur along an alternative reaction pathway that is not a reversal of the forward stepping pathway and that involves ATP hydrolysis rather than ATP synthesis, as proposed recently by Liepelt and Lipowsky \cite{Liepelt07PRL}. Based on structural considerations, however, we favor a backstepping cycle that is slightly different from the one proposed in Refs.\cite{Liepelt07PRL,Lipowsky08JSP}. We also argue that additional alternative pathways exist, although the ATP-induced backstepping cycle appears to be the dominant one under the commonly used experimental conditions.  

An important observation, which supports the existence of this alternative reaction route, is that the characteristic length scale $\delta (\approx 4 nm)$, obtained from the ratio of forward and backward step frequencies,  does not agree with the step size of kinesin as would be expected if backward steps are reversals of the forward steps. 
Furthermore, Yildiz \emph{et al.} recently shed light on the mechanical aspects of this issue by demonstrating the rather striking observation that as long as external loads supply enough mechanical energy,
kinesin can step both forward and backward even in the absence of ATP \cite{Yildiz08Cell}.

In the following section, we first review each step of the conventional single-cycle chemomechanics of kinesin movement in the absence of load by compiling the thermodynamic and kinetic information scattered over the literature. 
This section is useful not only for clarifying the terminology we use in this paper, but also for familiarizing the readership with the basic chemomechanics of kinesin motors before introducing the problematic issues of kinesin backstepping and providing our physical interpretation. 
In section III, we explain the various issues related to kinesin's backsteps from a large body of available data \cite{Cross05Nature,Nishiyama02NCB} and argue that kinesin's backsteps occur in at least one alternative route, different from the reverse cycle of single loop chemomechanics.  
Structures of molecular motors and their response to the external stress are complex and versatile, hence representing kinesin dynamics using a single cyclic reaction is not sufficient to grasp the essence of motor dynamics. 
Based on these arguments, we  will modify the conventional chemomechanics of kinesin and propose a plausible mechanism that can accommodate the puzzling data involving kinesin's backsteps (sections IV and V).

\section{Single-loop chemomechanics of kinesin}

\subsection{Thermodynamics, kinetics, and conformational changes } 

From a general perspective, biological systems maintain a nonequilibrium steady state condition through continuous supply of free energy and disposal or recycling of waste products. 
Thus, it is standard to represent processes of life in general by cyclic reactions \cite{Hill1989,Qian07ARPC}. Likewise, kinesin's motility has also been understood in terms of a single cyclic reaction powered by nucleotide chemistry.   
Following the binding of ATP to the nucleotide-free catalytic site of the kinesin motor domain, the catalysis of ATP (ATP hydrolysis and release of inorganic phosphate and ADP) induces a series of small and large conformational changes in the kinesin structure. 
The conformational change of kinesin rectifies the thermal fluctuation into 8 nm steps \cite{Yanagida05NatureChemBiol,Hyeon07PNAS2}.
The nucleotide chemistry at the two motor domains is coordinated in an out-of-phase manner, resulting in hand-over-hand stepping dynamics \cite{Block03Science,Schief04PNAS,Yildiz04Science}.     
One of the conventional reaction schemes for kinesin mechanics, suggested by studies based on a number of single-molecule and ensemble measurements, is illustrated in Fig. 1 
(see also Table I and II). In the following we summarize what is known about the transitions in this cycle, which will help understanding the physics of backstepping.

\subsubsection{ATP binding to the leading head : $D_{i-1}\cdot\phi_i\stackrel{+T}\longrightarrow D_{i-1}\cdot T_i$} 
This step involves a bi-molecular reaction of ATP binding to the nucleotide-free catalytic site of the leading head. 
We denote by $D_{i-1}\cdot\phi_i$ a kinesin configuration, in which the trailing head is in the ADP state and bound to the microtubule at the $i-1$-th tubulin binding site and the leading head is bound at the $i$-th tubulin binding site and contains no nucleotide in its catalytic site. Other motor states are denoted in an analogous fashion with $T$, $D$, $DP$, and $\phi$ representing the motor domains containing ATP, ADP, ADP$\cdot$P$_i$, and no nucleotide. 
The rate constant of this unit process depends on the ATP concentration $[T]$ and is given by 
$k_T=k_T^o[T]$. Under physiological conditions, the ATP concentration is $[T]\approx 1$ mM, thus $k_T\approx 2\times 10^3$ $sec^{-1}$ \cite{Moyer98BC}. 
If the ATP concentration is extremely low (e.g., 340 $nM$ in the experiment of ref.\ \cite{Yildiz04Science}), ATP binding becomes rate-limiting. 
In this case, the trailing head is already detached from the MT before the mechanical stepping occurs.  
Kinesin waits between steps in a two-head bound state at high ATP concentrations ($[T]=1$ $mM$) and in one-head bound state at low ATP concentrations ($[T]=1$ $\mu M$); this 
has been demonstrated directly in a recent FRET experiment by Mori \emph{et al.\ } \cite{Mori07Nature}.
To monitor the kinesin dynamics with the temporal resolution of an experimental setup, the ATP concentration is often adjusted to a low value \cite{Yildiz04Science}. 

\subsubsection{Mechanical stepping of the tethered head : $D_{i-1}\cdot T_i\rightarrow T_i\cdot D_{i+1}$} 
This step corresponds to the biggest change in time traces of kinesin displacement from bead experiments. 
ATP binding to kinesin's leading head induces a neck-linker docking transition (also termed the power stroke) in the motor domain. A disordered-to-ordered transition of the neck-linker transforms the underlying energy landscape of the ADP-containing trailing head, and biases (rectifies) the thermal diffusion of trailing head (tethered head) toward the next tubulin binding site \cite{Hyeon07PNAS2}.
Typically, the time scale of this stepping motion ($\lesssim 100$ $\mu s$) is extremely short ($\ll$ 1 \%) \cite{Nishiyama01NCB,Cross05Nature,Hyeon07PNAS2} compared to the entire reaction cycle that takes at least 10 $ms$ or longer.

\subsubsection{ADP release from the leading head : $T_i\cdot D_{i+1}\stackrel{-D}\longrightarrow T_i\cdot \phi_{i+1}$} 
Once the tethered head reaches the next tubulin binding site, the interaction between the MT and kinesin head containing ADP facilitates the dissociation of ADP from the catalytic site. 
There is a 100-fold difference in the ADP-dissociation rate from the catalytic site of the leading head ($k_{-D[L]}=75-100$ $s^{-1}$) compared to the dissociation from the trailing head ($k_{-D[T]}=1$ $s^{-1}$)  \cite{Ma97JBC,Uemura03NSB}.  The large difference between $k_{-D[L]}$ and $k_{-D[T]}$ is important for the coordination of the heads, because the affinity between the kinesin head and the MT changes from weak to strong upon ADP dissociation (see Table I). The small ADP-dissociation rate in the trailing head ($k_{-D[T]}\ll k_{-D[L]}$) thus ensures that ADP dissociates only after stepping.  It is also noteworthy that $k_{-D[T]}=1$ $s^{-1}$ is 100-fold slower than the complete kinesin cycle under saturating ATP condition. 
Therefore, it is very unlikely, \emph{for normal physiological concentrations of nucleotides and in the absence of opposing force}, that ADP dissociation occurs at the trailing head position.  
$k_{-D[L]}$ is comparable to the hydrolysis rate ($k_h>100\pm 30$ $s^{-1}$ \cite{Gilbert94Biochem}), thus ADP dissociation from leading head can become one of the rate limiting steps of the kinesin cycle. 
 
However, an opposing force, exerted onto the kinesin molecule, can bias the diffusion search space for the ADP containing tethered head to the backward direction. After a prolonged time of interaction with the MT due to the opposing force, ADP may be released even from the trailing head position. 
Once this occurs the strong affinity between the nucleotide-free kinesin head and MT can arrest the position of originally tethered head backward.   
The effect of opposing force on the kinetics of ADP release is important to understand kinesin's backsteps. We will discuss this issue in detail later in this article.  

\subsubsection{ATP hydrolysis and subsequent release of $P_i$ from the trailing head : $T_i\cdot\phi_{i+1}\rightarrow DP_i\cdot\phi_{i+1}\stackrel{-P}\longrightarrow D_i\cdot\phi_{i+1}$} 
An ATP bound to the catalytic site of the kinesin motor domain is hydrolyzed in 10 $ms$, and inorganic phosphate ($P_i$) is subsequently released. 
During these transitions, kinesin is strongly bound to the MT (see Table I). 
It has been argued that when both heads are strongly bound to the MT surface, tension on the neck-linker of the leading head strains the leading head structure \cite{Hyeon07PNAS,BlockPNAS06}. The catalytic site, in particular, is perturbed from its native-like environment, which affects the nucleotide chemistry. Although the same tension is exerted to the neck-linker of the trailing head, the catalytic site environment is not perturbed as much as the one in the leading head 
\cite{Hyeon07PNAS}. 
In this way, ATP binding to the leading head is regulated by the backward tension on the neck-linker \cite{Hyeon07PNAS,BlockPNAS06}. 
Recently, Yildiz \emph{et al.\ } have suggested, by showing the stepping dynamics of kinesin in the absence of ATP, that the forward tension exerted on the neck-linker of the trailing head plays a role in facilitating the detachment of the trailing head from the MT \cite{Yildiz08Cell}. It is thus possible that both forward and backward regulations by tension, which are not mutually exclusive, have distinct roles in the coordination of the kinesin heads.\\

Fig. 2 shows a free energy profile along the kinesin cycle by compiling  
the data measured for the individual reaction steps. This profile succinctly summarizes how the net free energy potential bias obtained from ATP hydrolysis, $\Delta \mu=\Delta\mu^o+k_BT\log{([D]_{cell}[P]_{cell}/[T]_{cell})}=-12 k_BT+k_BT\log{(70\cdot 10^{-6}\times 1\cdot 10^{-3}/1\cdot 10^{-3})}\approx -21.6 k_BT$ at physiological condition ($[T]=[T]_{cell}=1mM$, $[D]=[D]_{cell}=70 \mu M$ $[P]=[P]_{cell}=1mM$), is partitioned into the individual parts of the reaction cycle; $\Delta \mu^o\approx -12$ $k_BT$ is the free energy for the hydrolysis of an ATP molecule under standard solution condition ($T=25^oC$, 1 atm).   
It is of interest to note that the mechanical work ($W$) generated out of the total ATP hydrolysis energy is $W\approx -11.1\sim -10.0$ $k_BT$ (see Fig.2 and its caption), which is comparable to the one required to stall the kinesin motion $W_{stall}=-f_{stall}\times \delta \approx -(5-7) pN\times 8 nm\approx (-14\sim -10)$ $k_BT$, i.e., $W\approx W_{stall}$.    
There have been a number of discussions about whether models based on the neck-linker-docking picture provide sufficient energy to drive the system to step \cite{Block07BJ}. Some argue that free energy difference before and after the stepping is almost negligible \cite{Rice03BJ}. 
Here in contrast, purely relying on the thermodynamic analysis from Fig.2 and Table 2, we argue that free energy bias before and after the mechanical stepping ($\Delta G_{mech}\approx W=-10$ $k_BT$) is sufficient to drive the mechanical stepping. 
It is noteworthy that the ratio $W/\Delta G_{tot}\approx (0.5-0.6)$ is impressively high, compared to the thermodynamic efficiency of macroscopic motors. This ratio is even higher for 
myosin V and $\mathrm{F_1}$-ATPase \cite{Oster00JBB}. 
\\

\subsection{Michaelis-Menten representation of the kinesin cycle} 
From an enzymatic point of view, kinesins are ATPases that catalyze ATP hydrolysis. 
It has been shown that each mechanical step of kinesin results from the hydrolysis of a single ATP molecule, i.e., there is a strong coupling between ATP turnover and kinesin stepping with one molecule of ATP hydrolyzed for each 8 nm step along the MT \cite{Schnitzer97Nature}.
Thus, it is natural to describe kinesin's motility using Michaelis-Menten (MM) kinetics, which, in the absence of external load, indeed provides an accurate description the dependence of kinesin's ATPase rate as well as the dependence of motor velocity on the ATP concentration. 
Within the Michaelis-Menten model, the motor velocity of kinesin is given by   
\begin{equation}
  V=\delta \frac{k_{cat}[ATP]}{K_M+[ATP]}.
  \label{eqn:MM}
\end{equation}
where $\delta =8$ nm is the step size of the kinesin. 
The transport velocity saturates for $[ATP]\gtrsim 1$ $mM$ to $V_{max}=\delta k_{cat}\approx 8 nm/10 ms$ \cite{Visscher99Nature}, and a typical experimental value for the Michaelis constant is $K_M\gtrsim 50 \mu M$ \cite{Visscher99Nature,Moyer98BC}.  
The parameters $k_{cat}$ and $K_M$ can be expressed in terms of the more elementary rate constants of the cycle shown in Fig.\ 1A as 
$k_{cat}^{-1}=k_S^{-1}+k_{-D}^{-1}+k_h^{-1}+k_{-P}^{-1}$ and 
$K_M=k_{cat}/k_T\times\left(1+k_{-T}/k_S\right)$ (see Appendix I).

\subsection{Incorporation of the effect of external loads in the kinesin cycle} 
Much of our current knowledge about kinesin is due to single molecule experiments, which were developed approximately at the time of the discovery of kinesin \cite{Brady85Nature,Vale85Cell}, so that kinesin was one of the first molecules to be studied by single-molecule techniques \cite{Howard89Nature,Block90Nature}.
In addition to providing a direct visualization of the stepping of single kinesin motors \cite{Block94Cell,Vale96Nature,Yildiz04Science}, singe molecule experiments have also opened up another dimension to probe kinesin's behavior, namely by exerting mechanical forces that either oppose or assist the movement of the motor. 
A number of experiments have shown that external loads affect the nucleotide chemistry in the catalytic site \cite{Uemura03NSB,BlockPNAS06} as well as kinesin's motility. Thus, the force-velocity-ATP relationship has been employed to decipher the underlying mechanism of kinesin motility  
\cite{Block94Cell,Meyhofer95PNAS,Kojima97BJ,Visscher99Nature,Cross05Nature,Fisher01PNAS,Fisher05PNAS,Kolomeisky07ARPC,Tsygankov07PNAS}. 

The effect of load can be incorporated into the MM model by making the parameters $k_{cat}$ and $K_M$ force-dependent \cite{Schnitzer00NCB}. As long as the magnitude of load is small, MM kinetics fits the motility data well \cite{Visscher99Nature}.    
However, the validity of the conventional MM model to explain the kinesin motility becomes unclear when the external load approaches to the value of stall force ($\approx (5-7)$ pN). 
Importantly, an increasing backward load eventually stalls the motion of kinesin motion (i.e., $V=0$), and if increased further, induces backward stepping ($V<0$) \cite{Cross05Nature}. 
No modification of Eq.\ref{eqn:MM} can make the velocity completely zero or negative since the parameters, $k_{cat}$ and $K_M$, are both positive \cite{Fisher01PNAS}. 

One remedy for the failure of the MM model at large loads is to treat every reaction step within the kinesin cycle reversible (see Fig.\ 1B) \cite{Fisher01PNAS,Kolomeisky07ARPC}. Indeed, any chemical reaction is, in principle, reversible. In a reversible kinesin cycle, a resisting load can increase the probability of the reverse reactions and, thus, back-stepping. With this notion, one can suggest a fully reversible enzymatic reaction \cite{Fisher01PNAS}. 
When reversibility is imposed for every step of the kinesin cycle, as shown in Fig.\ 1B,
Eq.\ \ref{eqn:MM} is modified, which leads to \cite{Fisher01PNAS} (see Appendix II) 
\begin{eqnarray}
  V &=&\delta \frac{\prod_{i=1}^{N}{k^+_i[T]}-\prod_{i=1}^{N}{k_i^-[D][P]}}{\Sigma(\{k_i^{\pm}\})}\nonumber\\
  &=&\delta \frac{k^+_T[T]k_S^+k_{hyd}^+k_{-P}^+k_{-D}^+-k^-_Tk_S^-k_{hyd}^-k_{-P}^-[P]k_{-D}^-[D]}{\Sigma(\{k_i^{\pm}\})},
  \label{eqn:vel}
\end{eqnarray}
where 
the reaction constant $k_i^{\pm}$ ($i=T$, $S$, $-D$, $hyd$, and $-P$) for each step refers to ATP binding/dissociation ($k_T^{\pm}$), kinesin stepping/back-stepping ($k_S^{\pm}$), ATP hydrolysis/synthesis ($k_{hyd}^{\pm}$), phosphate release/rebinding ($k_{-P}^{\pm}$), and ADP release/rebinding ($k_{-D}^{\pm}$). The denominator $\Sigma(\{k_i^{\pm}\})$ is a lengthy but straightforward function of quartets of $k_i^{\pm}$. 
Here, $k_T^+$, $k_{-P}^-$, and $k_{-D}^-$ are bi-molecular rate constants while other rate constants are all uni-molecular. 
Because $k_i^{\pm}$ represent an elementary kinetic step, it is straightforward to incorporate the effect of force. 
According to the Bell model \cite{BellSCI78}, an external rearward force ($f$) affect the rate process by lowering the free energy barrier ($\Delta\Delta G_r^{\ddagger}=-f\delta^-$) for backward motion and raising the free energy barrier ($\Delta\Delta G_f^{\ddagger}=f\delta^+$) for forward motion, i.e., the rate for each elementary step is modified as  
\begin{equation}
  k_i^{\pm}\rightarrow  k_i^{\pm}\exp{(\pm f \delta_i^{\pm}/k_BT)}.
  \label{eqn:Bell}
\end{equation}
where $\delta_i^{\pm}$ is a distance from potential minimum to the barrier top in the one-dimensional reaction coordinate projected along the force direction which is also parallel to MT axis in our discussion. 
If any reaction step is irrelevant to the external load, one can set $\delta_i^{\pm}=0$ and make the rate constant force-independent. However, the total step of kinesin should be 8 nm for geometrical reason, which imposes the constraint $\sum_i\delta_i^++\sum_i\delta_i^-=\delta=8$ nm. 
When Eq.\ref{eqn:Bell} is inserted to Eq.\ref{eqn:vel} 
the velocity of kinesin in the presence of external resisting load $f$ reads 
\begin{equation}
  V(f) =\delta \frac{\left(\prod_{i=1}^{N}{k^+_i[T]}\right)e^{-f\sum_i\delta_i^+/k_BT}-\left(\prod_{i=1}^{N}{k_i^-[D][P]}\right)e^{f\sum_i\delta_i^-/k_BT}}{\Sigma(f)},
\end{equation}
which has the structure of $j(=V/\delta)=j_+-j_-$ (Appendix II). 
The force-dependent flux ratio  associated with the chemomechanical cycle
is obtained as $K=j_+/j_-$ (see Appendix II and see also Appendix III where the local flux ratio is discussed), which leads to 
\begin{equation}
  K=K^oe^{-f\delta/k_BT}=\left(K_{eq}\frac{[T]}{[D][P]}\right)e^{-f\delta/k_BT}.
\label{eqn:K}
\end{equation}
This flux ratio corresponds to a chemical potential 
\begin{eqnarray}
  \Delta \mu&=&\Delta \mu_{eq}+k_BT\log{\left(\frac{[D][P]}{[T]}\right)}+f\delta \nonumber\\
  &=& k_BT\log{\left[\left(\frac{[D][P]}{[T]}\right)/\left(\frac{[D]_{eq}[P]_{eq}}{[T]_{eq}}\right)\right]}+f\delta 
\end{eqnarray}
where 
\begin{eqnarray}
  \Delta\mu_{eq}&\equiv&-k_BT\log{K_{eq}}=-k_BT\log{K_T^{eq}K_{hyd}^{eq}K_S^{eq}K_{-P}^{eq}K_{-D}^{eq}}\nonumber\\
  &=&-k_BT\log{\left(\frac{[D]_{eq}[P]_{eq}}{[T]_{eq}}\right)}. 
\end{eqnarray}
The stall force 
at physiological condition ($[T]=[T]_{cell}$, $[D]=[D]_{cell}$, $[P]=[P]_{cell}$) 
can be calculated by setting the flux ratio to one (or $\Delta \mu=0$) for the reaction cycle, which leads to 
\begin{equation}
  f_{stall}\delta=-\Delta\mu_{eq}-k_BT\log{\left(\frac{[D]_{cell}[P]_{cell}}{[T]_{cell}}\right)}.
  \label{eqn:stall}
\end{equation}
We note that according to this expression the stall force depends on the concentrations of ATP and its hydrolysis products. 
For physiological conditions, however, the stall force is estimated to be 
$f_{satll}=1/8.2nm\times\{12k_BT-k_BT\log{((1\cdot 10^{-3})\times (70\cdot 10^{-6})/(1\cdot 10^{-3}))}\}\times 4.1(pN\cdot nm/k_BT))\approx 11$ pN, which is too large in comparison to the experimentally measured values ($5-7$ pN). 
Also, several experiments have shown that the stall force is almost insensitive to the nucleotide concentration \cite{Cross05Nature,Nishiyama02NCB,Meyhofer95PNAS}, but Eq.\ref{eqn:stall} shows a dependence of $f_{stall}$ on the concentrations. 
In the section IV, we will propose modifications of the conventional chemomechanics beyond a single cycle in which the stall force becomes ATP-independent after more specifically addressing the issue of kinesin backstepping.

The flux ratio can be determined experimentally by counting the occurrence of forward and backward steps. A note of caution is, however, appropriate here. For transitions that are not cyclic, $K$ is simply given by the ratio of the forward and backward stepping frequencies. In that case, $K$ corresponds to the force-dependent equilibrium constant of the reaction.
This has been nicely demonstrated at the single molecule level for the reversible folding and unfolding of an RNA hairpin under constant tension \cite{Bustamante01Sci}. In that case, the RNA hops between two conformational states in a discrete fashion. Bustamante and coworkers have measured the force-dependent equilibrium constant as given by the ratio of the dwell times at folded and unfolded states and have shown that the extrapolated value at zero force agrees well with the equilibrium constant from ensemble measurement. Furthermore, the characteristic distance, $R$, of the exponential force dependence, $K^{RNA}_{SM}(f)\sim e^{fR/k_BT}$, was found to be consistent with the difference in the end-to-end distances between unfolded and folded hairpins \cite{Bustamante01Sci}.
For kinesin, which operates in a cycle, it is less clear whether the measured ratio of forward and backward stepping frequencies measures the flux ratio $K$ calculated above. This would be true if the all steps correspond to completed forward or backward cycles. In that case, $\delta$ corresponds to the step size. In general, not all steps correspond to complete cycles. Instead of the global flux ratio $K$, one can also consider the local flux ratio through the mechanical stepping transition. We show in Appendix III that, for forces around the stall force, this quantity exhibits the same force-dependence as the global flux ratio, so we will not dwell on the difference between the two quantities here. Experimentally, an exponential force-dependence is seen for forces ranging from about 2 pN to 10 pN, but with a characteristic lengths scale of $\delta =(3-4)$ nm \cite{Cross05Nature,Nishiyama02NCB}.

\section{The puzzles in backward steps of kinesins} 
In the absence of a resisting load, backward steps of kinesin are quite rare.
 With increasing opposing force, however, backsteps become more and more frequent \cite{Nishiyama02NCB,Cross05Nature}. If the force exceeds kinesin's stall force, reaching super-stall force value ($\sim 14$ pN), kinesin has been shown to move ``slowly'' backward by performing processive backsteps \cite{Cross05Nature}. 
Within single cycle models for the stepping of kinesin, backward steps are explained by the reversed chemomechanical cycle of forward steps. This explanation implies in particular, that ATP is synthesized when kinesin moves backward. Such behavior has indeed been shown for the rotary motor F$_1$ ATPase, which in the absence of force also hydrolyzes ATP, but synthesizes ATP when mechanically driven backward \cite{Itoh04Nature}.
For kinesin, however, a number of observations suggest that this picture is unlikely and that backward steps are not forward steps executed in reverse, i.e., there is at least a fraction of backward steps that proceed through an alternative reaction pathway.

\subsection{Backsteps in the absence of force}

The flux ratio ($K_{SM}^o=N_{forward}/N_{backward}=802$) under zero load extracted from SM experiment \cite{Cross05Nature} suggests that although backward steps are rare in the absence of load, they occur more often than what one might expect based on a single-cycle model from the chemical equilibrium constant of ATP hydrolysis ($K^o_{cycle}=e^{-\Delta\mu/k_BT}=e^{21.6}\approx 2.4\times 10^9$) or mechanical equilibrium constant between $k_S^+$ and $k_S^-$ ($K^o_{mech}=k_S^+/k_S^-\approx 10^5$). 
One may explain this discrepancy by proposing that a backstep is induced by ADP rebinding under backward load; 
the observed (mechanical) backsteps does not necessarily correspond to full reverse cycle involving ATP synthesis events.
It is also plausible that the individual backsteps, preceded and followed by forward steps (as well as the first or the last step in a series of consecutive backsteps), in general correspond to partial chemical cycles \cite{Tsygankov07PNAS}.

Furthermore, according to Schief \emph{et. al.} \cite{Schief04PNAS}, high ADP concentration slows down the movement of kinesin, but this effect can be rescued \emph{almost} completely (but not entirely) by increasing the ATP concentration \cite{Schief04PNAS}. This result indicates that the main effect of a high ADP concentration is a competitive inhibition of ATP binding \cite{Schief04PNAS}. 
The rest of the effect of a high ADP concentration on the motor velocity, which cannot be masked by increasing the ATP concentration \cite{Schief04PNAS} 
may indeed be a signature of backsteps occurring through the reversed forward cycle, i.e. of backward steps stimulated by the high ADP concentration.

%

\subsection{ATP hydrolysis at large forces and ATP independence of the stall force}

Within a single-cycle model, a motor operating at the stall force makes forward and backward transitions within the cycle with equal probabilities. ATP hydrolysis and ATP synthesis occur with the same rates and no net ATP hydrolysis should occur. In particular, it has been pointed out that stepping and ATP hydrolysis should therefore occur at very low rates under typical experimental conditions, where the ADP concentration and thus the ATP synthesis rate are small \cite{Liepelt07PRL}. The experiment of Carter \& Cross \cite{Cross05Nature} suggests, however, that ATP hydrolysis still occurs under stall conditions, although this has not been directly measured. 

Related to the last point is the observation that the stall force appears to be the same ($\approx$ 7 pN) for ATP concentrations of 10 $\mu$M and 1 mM \cite{Cross05Nature}. Similarly, in observations by other labs, the stall force exhibits at most a very weak dependence on the ATP concentration \cite{Nishiyama02NCB,Meyhofer95PNAS}. Thermodynamic equilibrium, which in a single-cycle model is attained at the stall force, however, requires the stall force to depend on the ATP concentration as shown by Eq.\ref{eqn:stall}. The concentrations of the products ADP and P$_i$ were not controlled in the experiments of Carter and Cross. 
If we assume that they are similar in both cases, the 100-fold difference in ATP concentration should lead to a 4.6-fold difference in stall force. Unless changes in ADP and/or P$_i$ concentrations exactly compensate the change in ATP concentration between the two experiments, the lack of ATP dependence of the stall force provides a strong indication that stalling of kinesin does not correspond to thermodynamic equilibrium as required within a single cycle model. 
Furthermore the same study shows that the velocity of processive backward stepping at forces above the stall force depends on the ATP concentration and that backward stepping is faster for higher ATP concentrations \cite{Cross05Nature}. Within a single-cycle model one would expect that ATP tends to slow down backward stepping by driving the chemomechanical cycle in the forward direction, and that ADP rather than ATP would be expected to stimulate backsteps. As mentioned above, however, the stimulation of backward steps by high ADP concentration appears to be weak  even in the absence of force.

\subsection{Backward steps under load}

Carter and Cross \cite{Cross05Nature} also measured the frequency of backsteps under varying force, which allows to determine the  force-dependent flux ratio $K_{SM}(f)$ as the ratio of the of number of forward and backward steps. 
An exponential force dependence is expected with a characteristic length scale $\delta$ given by the step size of kinesin regardless of how one defines this quantity
(see the discussion in section II-C and Appendix III).  

The flux ratio obtained in the experiments by Carter and Cross is indeed given by an exponential force dependence, 
$K_{SM}(f)=K_{SM}^o\exp{(-f\delta/k_BT)}=802\times \exp{(-0.95\times f)}$, but the length scale of the exponent, $\delta\approx$ 4 nm, does not match with the step size ($\approx 8$ nm) of kinesin.
Likewise, the corresponding result of Nishiyama et al.\ has a length scale of $\delta\approx$ 3 nm, again clearly smaller than the step size. Both experiments show this ``incorrect'' length scale over a wide range of forces and in particular for forces around the stall force, where stepping in both directions is sufficiently frequent to obtain reliable data. It is thus very unlikely that the small characteristic length is a signature of the difference between the local and global flux ratios as discussed in Appendix III, but rather indicates a more fundamental difficulty of a single-cycle model to represent backward steps. 

\subsection{Backstepping in the absence of nucleotides}

Finally, a recent report by Yildiz \emph{et al.\ } shows directly that kinesin can step backward processively in the complete absence of nucleotides if a sufficiently strong force is applied \cite{Yildiz08Cell}. 
Likewise, forward motion can be obtained in the absence of nucleotides with a sufficiently strong assisting force, that pulls the motor forward. In both cases, the motor trajectories consist of 8 nm steps in a hand-over-hand fashion and are very similar to trajectories obtained for the usual conditions where movement is driven by ATP hydrolysis. These result are rather striking, because essentially in all models for the dynamics of kinesin, it has been taken for granted that ATP is required for motility (see also Eqs.\ref{eqn:MM} and \ref{eqn:vel}). Forced stepping in both directions was stimulated by the presence of ADP, which produces a weak interaction between kinesin and MT, and inhibited by the non-hydrolysable ATP analog AMP-PNP compared to the nuceotide-free case.  
Yildiz \emph{et al.} used these results to show that the coordinated hand-over-hand motion of kinesin is due to the alternated action of intramolecular strain in the neck-linker region of the molecule, but in addition these observations directly show that stepping of kinesin, both forward and backward, can occur through multiple pathways and provides further evidence that backward steps do not require ATP synthesis.  \\

Taken together the observations summarized above strongly suggest that backward steps are not reversed forward steps and that kinesin can follow other reaction pathways in addition to the chemomechanical cycle that drives forward stepping. Furthermore, the existence of backsteps under different nucleotide conditions as demonstrated by Yildiz \emph{et al.} \cite{Yildiz08Cell} indicates the existence of several alternative pathways rather than a single one, which may, however, be used with different probabilities, depending on the experimental conditions. In the next section we will discuss how the chemomechanical cycle model of kinesin can be extended to account for these alternatives. We note that the existence of alternative pathways has been proposed several times in the kinesin literature, based on different experimental observations \cite{Nishiyama02NCB,Cross05Nature,Liepelt07PRL} and that a systematic modeling study by Liepelt and Lipowksy  has described the dynamics of kinesin by two different cycles for forward and backward steps has, based on a non-equilibrium thermodynamics analysis \cite{Liepelt07PRL,Lipowsky08JSP}, but it seems that this idea has not gained general acceptance in the field yet. 

We also emphasize that the existence of alternative pathways for backward steps does not completely exclude the possibility of backwards steps driven by \emph{ATP synthesis}. Indeed, Hackney has shown that ATP can be synthesized by kinesin \cite{Hackney05PNAS}. His experiment however addressed only ATP synthesis from free P$_{\rm i}$ and ADP still bound to the motor head. It therefore demonstrates only that the hydrolysis transition can be reversed and not that the full cycle is reversed. Hence, it is not clear whether the observed ATP synthesis in the experiment is truly coupled to backward stepping. Since ATP hydrolysis is the step that is most likely to be practically irreversible, this observation nevertheless suggests that the cycle may run in reverse. 
As mentioned above, although the main effect of high ADP concentrations is competitive inhibition of ATP binding, there is also a weak non-competitive part \cite{Schief04PNAS}, which can be attributed to reversed cycles and thus ATP synthesis.  

\section{Backstepping cycle}

Based on the observations summarized in the previous section, we now construct a minimal model for backward steps that proceed through an alternative cycle. This cycle is shown in Fig.\ 3A. It is constructed based on two major constraints: (i) Since backward stepping is dependent on the ATP concentration \cite{Cross05Nature}, backward steps should branch off the main (forward) cycle after ATP binding, but before the forward stepping transition (i.e. $D_{i-1}\cdot T_i\rightarrow T_i\cdot D_{i+1}$). (ii) The mechanical backward stepping transition should occur when the leading head is in the ADP-bound state with weak MT affinity (see Table I).
We note that the resulting description of backsteps as described below is quite similar to the one of Ref.\ \cite{Liepelt07PRL} in that the backsteps occur through an alternative cycle associated with ATP hydrolysis rather than ATP synthesis. However, the two descriptions differ in the specific role ATP plays in inducing the backward steps as discussed below.

The two constraints lead us to the following scenario for backward steps under large loads: 
In the absence of opposing force, forward stepping of kinesin is triggered upon ATP binding to the leading head. 
The neck-linker changes from a disordered to an ordered state by zippering onto the neck-linker binding motif ($\beta11$ motif), 
thus transforming the underlying energy landscape for the tethered head to undergo forward-biased diffusion \cite{Hyeon07PNAS2}. 
However, under a sufficiently large backward load, the conformational transition of the neck-linker is prevented or at least the diffusion of tethered head is biased in the backward direction. The search space for tethered head is only restricted to the rear space of the MT-bound kinesin head \cite{Hyeon07PNAS2}. 
Although the dissociation kinetics of ADP from kinesin's catalytic site is slower in the rear head ($k_{-D[T]}\ll k_{-D[L]}$),  ADP can eventually, after $k^{-1}_{-D[T]}\sim 1$ sec, dissociate from the rear head position. 
Once ADP dissociation occurs, the nucleotide free rear head binds strongly to the MTs. The ATP contained in the leading head undergoes hydrolysis and 
once ADP state is reached the kinesin head turns into a weakly bound state. 
After that, the strong tension exerted through the neck-linker enhances the detachment of the leading head from the MT, resulting in the backward step (see Fig.3A) \cite{Yildiz08Cell}. At high ATP concentration and under large backward force, 
this scenario for backward stepping can be repeated in a cyclic fashion. Since this backstepping scenario works in the presence of ATP, we denote it $\mathcal{B}_T$. An important difference of the backward stepping cycle $\mathcal{B}_T$  compared to the forward stepping cycle is that ADP dissociation occurs from the trailing head. As mentioned this is a slow process, that happens on a time scale of $\sim 1$ sec, which provides a mechanistic explanation for why kinesin walks backwards quite slowly \cite{Cross05Nature} compared to its velocity in processive forward stepping.  

As mentioned above, the difference between the cycle described here and the backstepping cycle in the model of Liepelt and Lipowsky \cite{Liepelt07PRL} lies in different role of ATP in inducing the backstep. 
In the model of Liepelt and Lipowsky, the backstep directly follows upon ATP binding to the trailing head (see the transitions represented using $4\rightarrow 5\rightarrow 2$ in Fig.1(c) of Ref.\cite{Liepelt07PRL}), while in our model, backsteps ($\mathcal{B}_T$) are realized only after a series of processes (ATP binding, hydrolysis, and $P_i$ release) at the leading head (see Fig.3A). In principle, both reactions are possible, but we favor the second one for the following reason: (i) Once the leading head becomes an ADP state with weak binding affinity to the MT, we expect that an external load can easily pull the head backward by detaching it from the MT. (ii) Furthermore, ATP binding to the empty trailing head ($\phi_i\cdot D_{i+1}$) readily produces a neck-linker zippered $T_{\rm i}\cdot D_{\rm i+1}$ state (see Table I), which demands an unzipping of neck-linker and makes the backward stepping much harder to be realized.

Under even larger forces, backward steps can also be realized without nucleotides as demonstrated by Yildiz \emph{et al.} \cite{Yildiz08Cell}, which provides another pathway for backsteps. The role of the external load in this case is to exert tension on the neck-linker, so that the kinesin heads can be detached in an alternating fashion from the MTs. Since tension is always applied to the leading head through the neck-linker and the binding affinity of the nucleotide-free kinesin state is strong, the velocity of backstepping by this mechanism is lower than in the pathway $\mathcal{B}_T$ described above. This second pathway for backward steps, which we denote $\mathcal{B}_{\phi}$, is shown in Fig.3B.

According to our model, the role of ATP binding in backward stepping is not related to the direct triggering of a power stroke. Rather, ATP is employed to produce an ADP containing low affinity state through its hydrolysis. This picture is supported by the observation of Yildiz \emph{et al.} that backward stepping under force in the absence of ATP can be stimulated by ADP, but requires higher force in the presence of the non-hydrolyzable ATP analogue AMP-PNP.

All three cycles are combined in the reaction scheme shown in Fig.4. We note that there are multiple ways in which the no-nucleotide backstepping cycle can be reached from the ATP-dependent backstepping cycle. Likewise, escape from this cycle can happen through multiple routes ($\mathcal{E}$).

\section{Expansion of the chemomechanical cycle of kinesins}

\subsection{Two-cycle model with ATP-dependent backsteps ($\mathcal{F}^+\oplus\mathcal{B}_{T}^+$)} 
We now consider some consequences of the existence of multiple cycles by expanding the single cycle model in several steps. We start with the simplest possible case, and assume that all backward steps follow the ATP-dependent pathway and that backward steps never correspond to ATP synthesis. This model thus consists of the forward cycle $\mathcal{F}^+$ and the backward cycle $\mathcal{B}_{T}^+$, which are both taken to be irreversible. It can be viewed as a branched Michaelis-Menten model with two alternative irreversible pathways that separate after (reversible) ATP binding. 

The flux $j$ in this system, and thus the motor velocity $V=j\delta$ is calculated using 
\begin{equation}
j=k^+_S(f)P^{ss}_{D_{i-1}\cdot T_i}-k_{\mathcal{B}^+_T}^{eff}(f)P^{ss}_{D_{i-1}\cdot T_i}.
\end{equation}
where $k^{eff}_{\mathcal{B}^+_T}=(1/k_{-D[T]}^++1/k_{hyd}^++1/k_{-P}^++1/k_S^-)^{-1}$ is an effective rate for the backward step  (see Appendix II for more detail). 
The force-dependent flux ratio for this model is found to be given by  
\begin{equation}
  K_{\mathcal{F}^+\oplus\mathcal{B}^+_T}(f)=\frac{j_+}{j_-}=\frac{k_S^+}{k_{\mathcal{B}^+_T}^{eff}}\exp{\left(-\frac{f(\delta_{\mathcal{F}^+}+\delta_{\mathcal{B}^+_T})}{k_BT}\right)}.
\label{eqn:model1}
\end{equation}
Here, the characteristic lengths $\delta_{\mathcal{F}^+}$ and $\delta_{\mathcal{B}^+}$ correspond to the distances between the positions of the transition state and the local free energy minimum along the forward and backward cycle ($\mathcal{F}^+$ and $\mathcal{B}_T^+$), respectively. We denote the corresponding distances for the reversed cycles (which are not included in the simple irreversible two-cycle model) by $\delta_{\mathcal{F}_-}$ and $\delta_{\mathcal{B}_-}$. 
For any single-loop cycle these distances have to satisfy the thermodynamic constraint $\delta=\delta^++\delta^-=8$ nm,  i.e., $\delta_{\mathcal{F}^+}+\delta_{\mathcal{F}^-}=\delta_{\mathcal{B}^+_T}+\delta_{\mathcal{B}^-_T}=8$. However, there is no such requirement for  $\delta_{\mathcal{F}^+}+\delta_{\mathcal{B}^+_T}$. 
If both the transition barriers for $\mathcal{F}^+$ and $\mathcal{B}_T^+$ are located close to the $[D_{i-1}\cdot T_i]$ state, it is therefore possible that $\delta_{\mathcal{F}^+}+\delta_{\mathcal{B}^+_T}\lesssim 4$ nm as observed experimentally \cite{Nishiyama02NCB,Cross05Nature}.

Furthermore, the equilibrium constant at zero force is $K^o_{eq}\equiv k_S^+/k_{\mathcal{B}^+_T}^{eff}$, which is neither equal to $K_{cycle}$ nor to $K_{mech}$ for $\mathcal{F}^{\pm}$ cycle. The stall force is obtained as $f_{stall}=k_BT/(\delta_{\mathcal{F}^+}+\delta_{\mathcal{B}^+_T})\times \log{(k_S^+/k^{eff}_{\mathcal{B}^+_T})}$. It is noteworthy that both the flux ratio $K_{\mathcal{F}^+\oplus\mathcal{B}^+_T}(f)$ and the stall force do not depend on the ATP concentration, in accord with the data of Carter and Cross.   

This simplest possible two-cycle expansion of the conventional chemomechanics of kinesin thus captures most of the experimental results for backstepping \cite{Nishiyama02NCB,Cross05Nature}, which are mysterious within single-cycle models. Despite its simplicity and its limitations, it thus provides a clear idea of the origin of these peculiar results.\\ 

\subsection{Two-cycle model including ATP synthesis ($\mathcal{F}^+\oplus\mathcal{F}^-\oplus \mathcal{B}^+_T$)}
As discussed above, there is some evidence that kinesin can synthesize ATP, although this may be a rare event under typical experimental conditions. Adding an additional level of complexity, we now consider a two-cycle model consisting of a reversible forward stepping cycle ($\mathcal{F}^{\pm}$) combined with the ATP-induced backstep cycle $\mathcal{B}^+_T$. The backward stepping cycle is still taken to be irreversible. In this model, backward steps can occur along two different pathways, either induced by ATP binding and coupled to ATP hydrolysis as before or coupled to ATP synthesis. The total flux is then measured by  
\begin{equation}
j=k_S^+(f)P^{ss}_{D_{i-1}\cdot T_i}-k_S^-(f)P^{ss}_{T_i\cdot D_{i+1}}-k_{\mathcal{B}^+_T}^{eff}(f)P^{ss}_{D_{i-1}\cdot T_i}
\end{equation}
and this leads to a flux ratio ($K=j_+/j_-$) as follows 
\begin{eqnarray}
 K_{\mathcal{F}^+\oplus\mathcal{F}^-\oplus\mathcal{B}^+_T}(f)
=\frac{[T]k_T^+k_S^+k_{-D[L]}^+k_{hyd}^+k_{-P}^+e^{-f\delta_{\mathcal{F}^+}/k_BT}}{[D][P]k_T^-k_S^-k_{-D[L]}^-k_{hyd}^-k_{-P}^-e^{f\delta_{\mathcal{F}^-}/k_BT}+A(f)k_{\mathcal{B}_T^+}^{eff}e^{f\delta_{\mathcal{B}^+_T}/k_BT}}
\label{eqn:model2}
\end{eqnarray}
where $A(f)\equiv \{[T]k_T^+k_{hyd}^+k_{-P}^+(k_{-D[L]}^++k_S^-)+[T][D]k_T^+k_S^-k_{-D[L]}^-(k_{-P}^++k_{hyd}^-)+2[D][P]k_S^-k_{-D[L]}^-k_{hyd}^-k_{-P}^-\}(f)$. 

Several comments are in order here:
(i) If backstepping due to ATP synthesis is zero (i.e., $k_S^-=0$) Eq.\ref{eqn:model2} is reduced to Eq.\ref{eqn:model1}. 

(ii) When the limit of low ADP concentration, $[D]\rightarrow 0$, we obtain 
\begin{equation}
K_{\mathcal{F}^+\oplus\mathcal{F}^-\oplus\mathcal{B}^+_T}(f)\rightarrow 
\frac{k_S^+k^+_{-D[L]}}{(k_S^-+k^+_{-D[L]})k_{\mathcal{B}^+_T}^{eff}}\exp{\left(-\frac{f(\delta_{\mathcal{F}^+}+\delta_{\mathcal{B}^+_T})}{k_BT}\right)}.
\end{equation}
Similar to the simplest two-cycle model ($\mathcal{F}^+\oplus\mathcal{B}_T^+$), low ADP concentrations result in a linear relationship between $\log{K_{\mathcal{F}^+\oplus\mathcal{F}^-\oplus\mathcal{B}^+_T}}(f)$ and $f$ curve, as observed in Carter and Cross' experiment.  

(iii) For general conditions, where reaction constants associated with $\mathcal{F}^-$ cycle are not negligible, and the concentration of ADP and P$_\mathrm{i}$ may be high, a simple linear relation for the [$f$,$\log{K_{\mathcal{F}^+\oplus\mathcal{F}^-\oplus\mathcal{B}^+_T}}(f)$] plot will not hold.  
If backsteps are associated with more than two cycles, a nonlinearity manifests itself in the relation of Log(flux ratio) and $f$. 
\\

\subsection{Multiple-cycle model including nucleotide-free backstepping ($\mathcal{F}^+\oplus\mathcal{F}^-\oplus\mathcal{B}^+_T\oplus\mathcal{B}^+_{\phi}$)} 
By further combining the reversible single-loop cycle with the backstep cycles $\mathcal{B}_T^+$ and $\mathcal{B}_{\phi}^+$ one can suggest an expanded chemomechanical cycle as in Fig.4. 
Although an exact expression of the flux ratio  can be obtained by solving the master equation as in the above examples, the resulting expression is rather complicated with many reaction constants. The general form of  the flux ratio is given by  
\begin{equation}
  K_{\mathcal{F}^+\oplus\mathcal{F}^-\oplus \mathcal{B}^+_T\oplus\mathcal{B}^+_{\phi}}(f)=\frac{[T]\mathcal{C}_{\mathcal{F}^+}e^{-f\delta_{\mathcal{F}^+}/k_BT}}{[D][P]\mathcal{C}_{\mathcal{F}^-}e^{f\delta_{\mathcal{F}^-}/k_BT}+\mathcal{X}_{\mathcal{B}_T^+}(f)e^{f\delta_{\mathcal{B}_T^+}/k_BT}+\mathcal{X}_{\mathcal{B}_{\phi}^+}(f)e^{f\delta_{\mathcal{B}_{\phi}^+}/k_BT}}.
\label{eqn:full}
\end{equation}
The structure of $K_{\mathcal{F}^+\oplus\mathcal{F}^-\oplus \mathcal{B}^+_T\oplus\mathcal{B}^+_{\phi}}(f)$ suggests that to observe a linear relationship between $\log{K(f)}$ and $f$, only one of the backward cycle should contribute to the entire flux; otherwise, we expect deviations from the linear relation in the $[f,\log{K(f)}]$ plot. 

It is of interest to note that a very similar kinesin experiment by Visscher \emph{et al}. \cite{Visscher99Nature}, who used position clamp conditions to probe the stall force, exhibited a stronger ATP dependence of the stall force. It may be plausible that the position clamp condition produces a different mechanical response of kinesin compared to the one obtained under the fixed trap condition other labs \cite{Nishiyama02NCB,Meyhofer95PNAS,Cross05Nature} have used in their studies. In the position clamp condition \cite{Visscher99Nature}, kinesin possibly adopts a mixture of cycles responsible for backstepping where the contributions from $\mathcal{B}_{\phi}$ and/or $\mathcal{F}^-$ cycles for backstepping become more pronounced than the $\mathcal{B}_T$ cycle (see Eq.\ref{eqn:full}).   
We would also like to point out that the stall force values measured by Visscher et al. \cite{Visscher99Nature} (5-8 pN for 10 $\mu M-1$ mM ATP) are still smaller than 11 pN at 1mM-ATP, the latter of which we obtained for physiological concentrations of the nucleotides.  

Fig.5 shows the $K(f)$ data from Carter and Cross' experiment \cite{Cross05Nature} and its fits at two different ATP concentrations. 
To fit the data we modified Eq.\ref{eqn:full} as follows by assuming that one other cycle is involved with backstepping.  
\begin{equation}
K(f)=(802 \times e^{0.95\times f})\times \left(\frac{1}{1+K'e^{f\Delta /k_BT}}\right)
\end{equation}
where $\Delta\equiv \delta_{\mathcal{B}_{\phi}^+}-\delta_{\mathcal{B}_T^+}$ for instance ($-8<\Delta<8)$. 
The factor inside the parenthesis is a correction factor to Carter and Cross's original fit. 
However, with the parameters determined from the fits ($K'\approx 1$ and $\Delta<0$, see the caption to Fig.\ 5), the contribution from the correction factor is practically negligible at $f>1$ pN. 
A cycle whose activation relies more on the external force ($\mathcal{B}_{\phi}^+$ cycle for instance) is expected to have a smaller $\delta^+$ value, so that $\Delta<0$ will ensure that such a cycle contributes little to $K(f)$ value at higher force. 
We expect up to 50 \% deviation from the linear behavior when $f\rightarrow 0$ (see Fig.5). 
In the absence of external load, more than two routes will affect the flux ratio. 

\section{Concluding Remarks}

Based on the analysis of experimental data by Carter and Cross \cite{Cross05Nature}, we have argued that backward steps of kinesin
 do not occur through the reversal of a forward chemomechanical cycle, but rather occur through an alternative cycle that depends on ATP binding.
The role of ATP binding for the backward steps is not equivalent to the one in forward step, where ATP binding induces the conformational change in kinesin's structure. 
In backstepping under opposing load,  
ATP is employed to produce an ADP-bound state in the leading head to weaken its binding to the MT surface, so that the external force can easily detach the leading head and move it backward (see Fig.3A).
This simple extension of the conventional chemomechanics of kinesin, which leads to a model with two cyclic reaction pathways ($\mathcal{F}^+\oplus\mathcal{B}_T^+$), can account for most experimetal findings on backward steps. 

(i) The velocity of processive backward stepping under high force is small compared to the velocity of forward stepping. In the $\mathcal{B}_T^+$ cycle, ADP dissociates from the trailing head, which is about two orders of magnitude slower ($k_{-D[T]}\approx 1$ $s^{-1}$) than the ADP dissociation from the leading head ($k_{-D[L]}\approx 75-100$ $s^{-1}$). 
Our proposal for the mechanism of ADP dissociation from the trailing head in the ATP dependent backstep ($\mathcal{B}_T$ cycle) is also supported by the observation that dwell time is two orders of magnitude longer for backward step compared to the forward steps. 

(ii) In the two-cycle model both forward and backward step consume one ATP molecule per step, so the dependence of ATP in the forward and backward fluxes exactly cancel out both in the flux ratio and in the stall force. This provides a simple explanation why no (or at most a weak) ATP-dependence of the stall force is observed. It also agrees with the observation that the force-dependent flux ratio is essentially the same over the wide range of force values at both low and high ATP concentrations. 

(iii) Finally, this model provides a simple interpretation for the characteristic length scale of the flux ratio ($\delta_{\mathcal{F}^+}+\delta_{\mathcal{B}_T^+}\approx (3-4)$ nm \cite{Cross05Nature,Nishiyama02NCB}) and suggests that the transition barriers for both forward and backward cycle are located closely to the starting conformation of kinesin. 

While there is evidence for the existence of additional pathways such as forced stepping in the absence of nucleotides ($\mathcal{B}_{\phi}^+$), the ATP-independence and single-exponential behavior of the flux ratio indicates that the behavior of kinesin for typical experimental nucleotide concentrations and under load up to and around the stall force is dominated by only two cycles, the ATP-driven forward stepping cycle and the ATP-dependent backward stepping cycle ($\mathcal{F}^+\oplus\mathcal{B}_T^+$). 
Other pathway may become more important for high ADP concentrations or for even higher forces near super-stall. 

Even though we have focused on ``hand-over-hand'' backward steps in this article, 
8-nm backward ``slips'' with the same head rebinding is another plausible mechanism that we cannot totally exclude from the mechanism for backward steps \cite{Cross05Nature,Yildiz08Cell}.
Other forward step mechanisms as well as a futile cycle that do not lead to any step may also contribute to the pathway heterogeneity for kinesin dynamics. While the forward cycle shown in Fig.1 describes well the experimental data for zero force and weak opposing force, so that there seems to be no need for alternative pathways here, the forward stepping of kinesin under assisting loads is less well understood \cite{Block07BJ} and the force-velocity relations measured in different labs do not agree very well in this regime \cite{Coppin97PNAS,Block03PNAS,Cross05Nature}. 
It is again noteworthy to mention Yildiz {\it et al.}'s experiment that has demonstrated forward stepping without nucleotide under assisting loads \cite{Yildiz08Cell}.  
Finally, some pathway heterogeneity may even be possible within the normal forward cycle, but hard to resolve experimentally, if it only pertains to where in the cycle a non-rate-limiting step occurs. For example, it is possible that in some steps ATP hydrolysis in the rear head happens before ADP is released from the leading head. We note that such partitioning of reaction routes on the main forward stepping cycle has recently been proposed for Myosin V motors \cite{Baker04PNAS,Uemura04NSMB,Wu07Biochem}.

The complexity of biomolecular structures
makes it common to observe partitioning of dynamics into multiple routes, as exemplified in protein and RNA folding studies \cite{Scherer08PNAS,HyeonBC05,RussellPNAS02,Mickler07PNAS}. 
The partitioning ratio is a sensitive function of external conditions, so that different routes become dominant under different experimental conditions.  
For molecular motor systems in nonequilibrium steady state, coincidence of reaction routes for forward and backward cycle is not necessary.  
Despite the relative simplicity in comparison to other molecular motors, 
the architecture of kinesins and its ligand-dependent conformational changes coupled to the interactions with MTs are complex enough that the multifaceted response of kinesin lend itself under external loads.\\ 

{\bf Acknowledgements}
 
This work was supported by the Center for Theoretical Biological Physics sponsored by the National Science Foundation (NSF) (Grant PHY-0822283) with additional support from NSF Grant MCB-053906. C.H. acknowledges support from the Korea Sicence and Engineering Foundation grant funded by the Korean Government Ministry of Science and Technology (Grant R01-2008-000-10920-0) and the Korea Research Foundation Grant funded by Korean Government Grants (KRF-C00142 and KRF-C00180). S.K. was supported in part by a fellowship from Deutsche Forschungsgemeinschaft (Grants KL818/1-1 and 1-2).

\section*{Appendix I}
For an irreversible MM kinetics with multiple steps, 
\begin{equation}
  A_0+S\mathop{\rightleftharpoons}^{k_0}_{k_0'} A_1\stackrel{k_1}{\rightarrow} A_2\stackrel{k_2}{\rightarrow}\cdots\rightarrow A_N\stackrel{k_N}{\rightarrow} A_0+P,
\end{equation}
the formation rate of the product $P$ at steady state is given by 
\begin{equation}
  V=\frac{k_{cat}^{eff}[A]_{tot}[S]}{K_M^{eff}+[S]}
\end{equation}
where the mass balance law, $[A]_{tot}=\sum_{i=1}^N[A_i]$, is used and 
\begin{eqnarray}
  k_{cat}^{eff}&=&\left(\sum_{i=1}^N k_i^{-1}\right)^{-1}\nonumber\\
  K_M^{eff}&=&\frac{k_{cat}^{eff}}{k_1}\left(\frac{k_1+k_0'}{k_0}\right)
\end{eqnarray}

\section*{Appendix II}
For simplicity, when a cyclic reaction is made of three chemical species ($N=3$), of which the total probability is normalized to a unity ($P_{E_1}+P_{E_2}+P_{E_3}=1$), 
\begin{equation}
  E_1\mathop{\rightleftharpoons}^{k_1}_{k_{-1}} E_2\mathop{\rightleftharpoons}^{k_2}_{k_{-2}} E_3\mathop{\rightleftharpoons}^{k_3}_{k_{-3}} E_1 
\end{equation}
a steady state condition of master equation gives the probability of each chemical species as 
\begin{eqnarray}
P^{ss}_{E_1}&=&\frac{k_2^+k_3^++k_3^+k_1^-+k_1^-k_2^-}{\Sigma(\{k^{\pm}_i\})}
\nonumber\\
P^{ss}_{E_2}&=&\frac{k_3^+k_1^++k_1^+k_2^-+k_2^-k_3^-}{\Sigma(\{k^{\pm}_i\})}
\nonumber\\
P^{ss}_{E_3}&=&\frac{k_1^+k_2^++k_2^+k_3^-+k_3^-k_1^-}{\Sigma(\{k^{\pm}_i\})}
\label{eqn:ss}
\end{eqnarray}
where $\Sigma(\{k^{\pm}_i\})=k_1^+k_2^++k_2^+k_3^++k_3^+k_1^++k_1^-k_2^-+k_2^-k_3^-+k_3^-k_1^-+k_1^-k_3^++k_1^+k_2^-+k_2^+k_3^-$. 

The definition of velocity $V_3\equiv k_i^+P^{ss}_{E_i}-k_i^-P^{ss}_{E_{i+1}}$ leads to  
\begin{equation}
  V_3=\frac{k_1^+k_2^+k_3^+-k_1^-k_2^-k_3^-}{k_1^+k_2^++k_2^+k_3^++k_3^+k_1^++k_1^-k_2^-+k_2^-k_3^-+k_3^-k_1^-+k_1^-k_3^++k_1^+k_2^-+k_2^+k_3^-}
\end{equation}
This tri-cyclic reversible reaction can be recast to the well-known reversible enzyme kinetics, $E+S\rightleftharpoons ES\rightleftharpoons EP\rightleftharpoons E+P$ \cite{Alberty63JBC,Brant63JACS} as
\begin{equation}
  V_3 = \frac{\frac{v_S[S]}{K_M^S}-\frac{v_P[P]}{K_M^P}}{1+\frac{[S]}{K_M^S}+\frac{[P]}{K_M^P}},
\end{equation}
where $v_S=\frac{k_2^+k_3^+}{k_2^++k_2^-+k_3^+}[E]_{tot}$, $v_P=\frac{k_1^-k_2^-}{k_1^-+k_2^++k_2^-}[E]_{tot}$, $K_M^S=\frac{k_1^+k_2^-+k_1^-k_3^++k_2^+k_3^+}{k_1^+(k_2^++k_2^-+k_3^+)}$, $K_M^P=\frac{k_1^+k_2^-+k_1^-k_3^++k_2^+k_3^+}{k_3^-(k_1^-+k_2^++k_2^-)}$ and $[E]_{tot}\equiv [E]+[ES]+[EP]$. 
The enzyme interconverts $S\rightleftharpoons P$ reversibly. 
The cyclic reaction is stalled or alternatively the microscopic reversibility is established for the cycle when $V_3=0$. This condition is accomplished when 
$v_S[S]_{eq}/K_M^S=v_P[P]_{eq}/K_M^P$, which leads to 
\begin{equation}
  K_{eq}\equiv\frac{[P]_{eq}}{[S]_{eq}}=\frac{v_S}{v_P}\frac{K_M^P}{K_M^S}=\frac{k_1^+k_2^+k_3^+}{k_1^-k_2^-k_3^-}=K_1^{eq}K_2^{eq}K_3^{eq}.
\label{eqn:Keq}
\end{equation}
where $K_i^{eq}\equiv k_i^+/k_i^-$ with $i=1,2,3$. 
If one defines positive and negative fluxes as $J_+\equiv \frac{v_S[S]/K_M^S}{1+[S]/K_M^S+[P]/K_M^P}$ and $J_-\equiv \frac{v_P[P]/K_M^P}{1+[S]/K_M^S+[P]/K_M^P}$ 
then $V_3\equiv J=J_+-J_-$ and 
$K\equiv J_+/J_-=K_{eq}\frac{[S]}{[P]}=\frac{[S]/[S]_{eq}}{[P]/[P]_{eq}}$. Thus, one can define a chemical potential for a cyclic reaction. 
\begin{equation}
  \Delta \mu\equiv-k_BT\log{K}=\Delta \mu_{eq}+k_BT\log{\frac{[P]}{[S]}}.
\end{equation}
If $\Delta \mu < 0$ then the reaction cycle proceeds in the plus direction, which is maintained as long as $[P]$ and $[S]$ are kept far from equilibrium such that $[S]\gg [P]$.  

\section*{Appendix III}
In addition to the mechanical steps realized by full cycle, it is also plausible to think about the local stepping dynamics of kinesins. For instance, a backstep can be made by ADP rebinding to the leading head, and repeated forward and backward reaction may occur in rapid succesion, which may not be resolved in the experimental timescale.  
In the bead experiments, it is almost impossible to distinguish whether the step is realized from a full cycle or the step is simply realized locally. 
Nevertheless, 
one can at least propose a definition of local flux ratio and compare it with the flux ratio of the full cycle.  

For simplicity, we use the cyclic reaction associated with three chemical species as in Appendix II to represent the local flux ratio between $E_1\rightleftharpoons E_2$ as follows. 
\begin{eqnarray}
K_{local}=\frac{k_1^+P_{E_1}^{ss}}{k_1^-P_{E_2}}&=&\frac{k_1^+k_2^+k_3^++k_1^+k_3^+k_1^-+k_1^+k_1^-k_2^-}{k_1^-k_2^-k_3^-+k_1^+k_3^+k_1^-+k_1^+k_1^-k_2^-}\nonumber\\
&=&K_{eq}\left[1+\underbrace{\frac{(k_1^-/k_2^+-k_1^+k_3^+/k_2^-k_3^-)+(k_1^-k_2^-/k^+_2k^+_3-k_1^+/k_3^-)}{1+k_1^+/k_3^-+k^+_1k_3^+/k^-_2k^-_3}}_{X}\right]
\label{eqn:Klocal}
\end{eqnarray} 
where $K_{eq}$ has the same definition with Eq.\ref{eqn:Keq}.
At stall condition $K_{local}=1$, one gets $k_1^+k_2^+k_3^+=k_1^-k_2^-k_3^-$, and the 
terms underbraced with $X$ vanishes. Thus, 
the local flux ratio and the full flux ratio give the same dependence of $f$. 
If $f\neq f_{stall}$, the force dependence of Eq.\ref{eqn:Klocal}, 
$K_{local}(f)=K_{eq}^oe^{-f\delta/k_BT}A(f)$ where $A(f)$ is a complicated function of $f$, 
is not exponential as in Eq.\ref{eqn:K}. In general,
$\log{K_{local}(f)}$ becomes linear in $f$ only near the stall force. In contrast, in Carter and Cross' experiment $\log{K(f)}$ vs $f$ is linear over a wide range of force values, 1 pN$\lesssim f\lesssim 10$ pN \cite{Cross05Nature}.   

\clearpage


\clearpage
\begin{table}
\begin{center}
\caption{\label{tab1}Binding affinity between monomeric kinesins and MTs, neck-linker conformation with different nucleotide states
}
\begin{tabular}{l|c|c}
\hline
&\footnote{All kinesin family members show a common pattern of dissociation constant with ligand $\mathrm{ADP>ADP\cdot AlF_4>AMPPNP>\phi}$ where $\mathrm{ADP\cdot AlF_4}$ and AMPPNP are the analogs of $\mathrm{ADP\cdot P_i}$ and ATP, respectively. The nucleotide-free state ($\phi$) binds most tightly to microtubules \cite{Cross00PTRSL}.} Binding affinity to MTs \cite{Cross00PTRSL} & Neck-linker state \cite{ValeNature99} \\ \hline
$K\cdot \phi$ & Strong & Unzippered (disordered) \\
$K\cdot T$  & Strong& Zippered (ordered) \\ 
$K\cdot DP$ & Strong& Zippered (ordered) \\ 
$K\cdot D$ & Weak & Unzippered (disordered) \\ \hline
\end{tabular}
\end{center}
\end{table}

\begin{table}
\begin{center}
\caption{\label{tab2}Rate and equilibrium constants for the microscopic steps along the cycle}
\begin{tabular}{|l||l|l|}
\hline
&Rate constant & Equilibrium constant\\ \hline\hline
\footnote{Bi-molecular rate constant for ATP binding to the nucleotide free kinesin head}$k_T^+(\equiv k_T)$ &  $k_T^o=2.0\pm 0.8$ $\mu M^{-1}s^{-1}$ \cite{Moyer98BC} &$K_T=k^o_T/k_{-T}=(35$ $\mu M)^{-1}$ \cite{Moyer98BC} \\
\footnote{Rate constant for ATP dissociation from the ATP bound kinesin head}$k_T^-(\equiv k_{-T})$ &$k_{-T}=71\pm 9 s^{-1}$ \cite{Moyer98BC} & \\ \hline
\footnote{Rate constant for kinesin stepping}$k_S^+(\equiv k_S)$ & $k_S\gtrsim (100$ $\mu s)^{-1}$& no data \\ 
\footnote{Rate constant for kinesin backstepping}$k_S^-(\equiv k_{-S})$ & & \\ \hline
\footnote{Rate constant for ADP dissociation}$k_{-D}^+(\equiv k_{-D})$ & $k_{-D[L]}=75-100$ $s^{-1}$\cite{Ma97JBC} & $K_{D[L]}=k_D^o/k_{-D}=5\times 10^4 M^{-1}$ \cite{Ma97JBC}\\
& $k_{-D[T]}=1$ $s^{-1}$\cite{Ma97JBC} & $K_{D[T]}=k^o_D/k_{-D}=5\times 10^6 M^{-1}$ \cite{Ma97JBC}\\
\footnote{Rate constant for ADP rebinding}$k_{-D}^-(\equiv k_D)$ & no data & \\ \hline
\footnote{Rate constant for hydrolysis at the catalytic site. The free energy change for ATP hydrolysis $\mathrm{ATP\leftrightharpoons ADP+P_i}$ under standard aqueous conditions ($aq$, 1 atm, 25$^oC$) is $\Delta\mu^o=-12$ $k_BT$.}$k_{h}^+(\equiv k_h)$ & $k_h> 100\pm 30$ $s^{-1}$ \cite{Gilbert94Biochem} &$K_h<39$. \\ 
\footnote{Rate constant for ATP synthesis}$k_{h}^-(\equiv k_{-h})$ & $k_{-h}=1.3$ $s^{-1}$ & \\ \hline
\footnote{Rate constant for Phosphate dissociation}$k_{-P}^+(\equiv k_{-P})$ & $k_{-P}=50$ $s^{-1}$ & no data\\
\footnote{Rate constant for Phosphate rebinding}$k_{-P}^-(\equiv k_{P})$ & no data & \\ \hline
\footnote{Rate constant for hydrolysis followed by phosphate release}$k_{h,-P}$ & $k_{h,-P}=100-300$ $s^{-1}$ &$K_{h,-P}=200s^{-1}/34M^{-1}s^{-1}=6M$ \cite{Hackney05PNAS} \\
& $k_{-h,P}=34$ $M^{-1}s^{-1}$ \cite{Hackney05PNAS} & \\
\hline

\hline
\end{tabular}
\end{center}
\end{table}

\clearpage
\begin{figure}[ht]
  \begin{center}
   \includegraphics[width=4in]{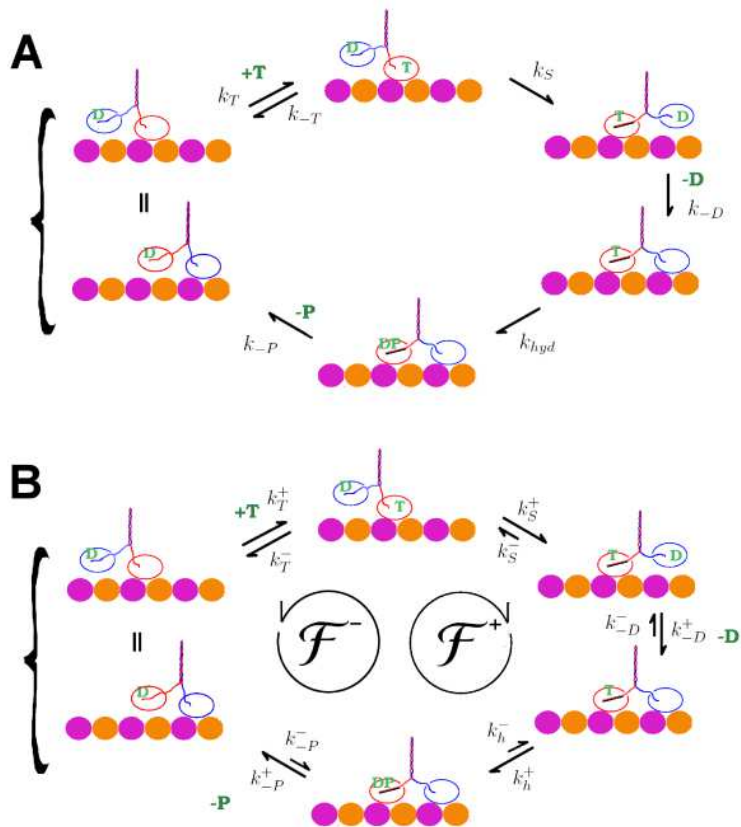}
   \caption{Conventional chemomechanical cycle for kinesin. A protofilament of a microtubule, with its minus end to the left and plus end to the right and the kinesin configuration at each reaction step are illustrated in the figures. {\bf A.} Irreversible, {\bf B.} Reversible case. Of the two motor heads, we designate the heads pointing into the plus and minus end directiontion as the \emph{leading} and \emph{trailing head}, respectively.  }
  \end{center}
\end{figure}

\begin{figure}[ht]
  \begin{center}
   \includegraphics[width=6in]{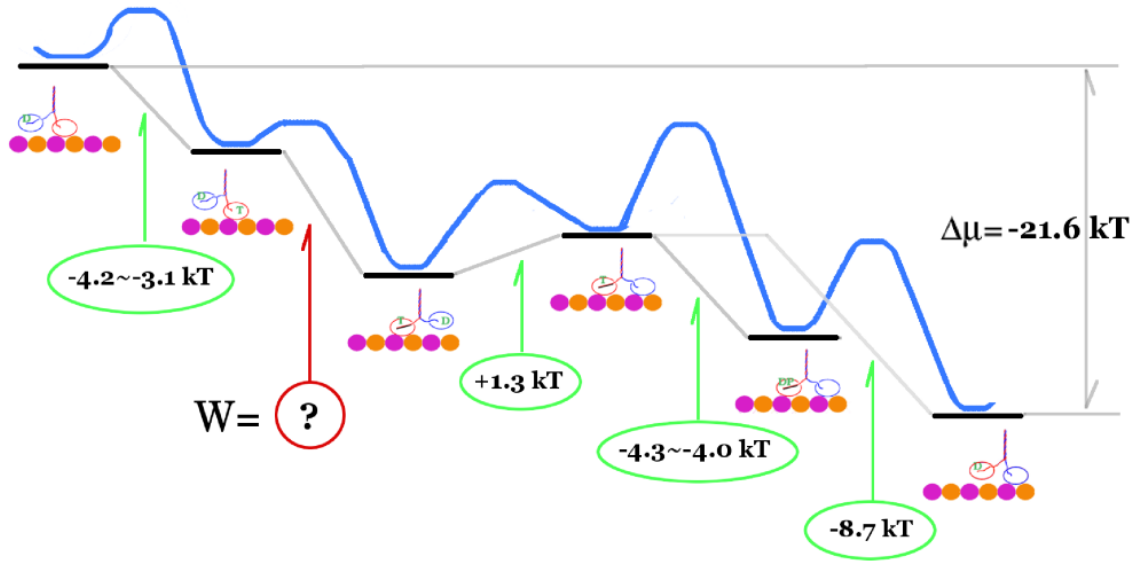}
   \caption{Free energy profile of the kinesin cycle. 
Standard equilibrium chemical thermodynamics and Table II are employed to calculate the free energy difference between the chemical states along the chemomechanical cycle shown in Fig.1B.  
Because of the lack of reliable thermodynamic (or kinetic) data, we treat the free energy change $W$ associated with the disorder-to-order transition of the neck-linker and the subsequent mechanical step as an unknown. We estimated $W\approx -11.0\sim -10.0$ $k_BT$ by subtracting the free energy differences from individual reaction steps from the net ATP hydrolysis free energy ($\Delta \mu$).}
  \end{center}
\end{figure}

\begin{figure}[ht]
  \begin{center}
   \includegraphics[width=4in]{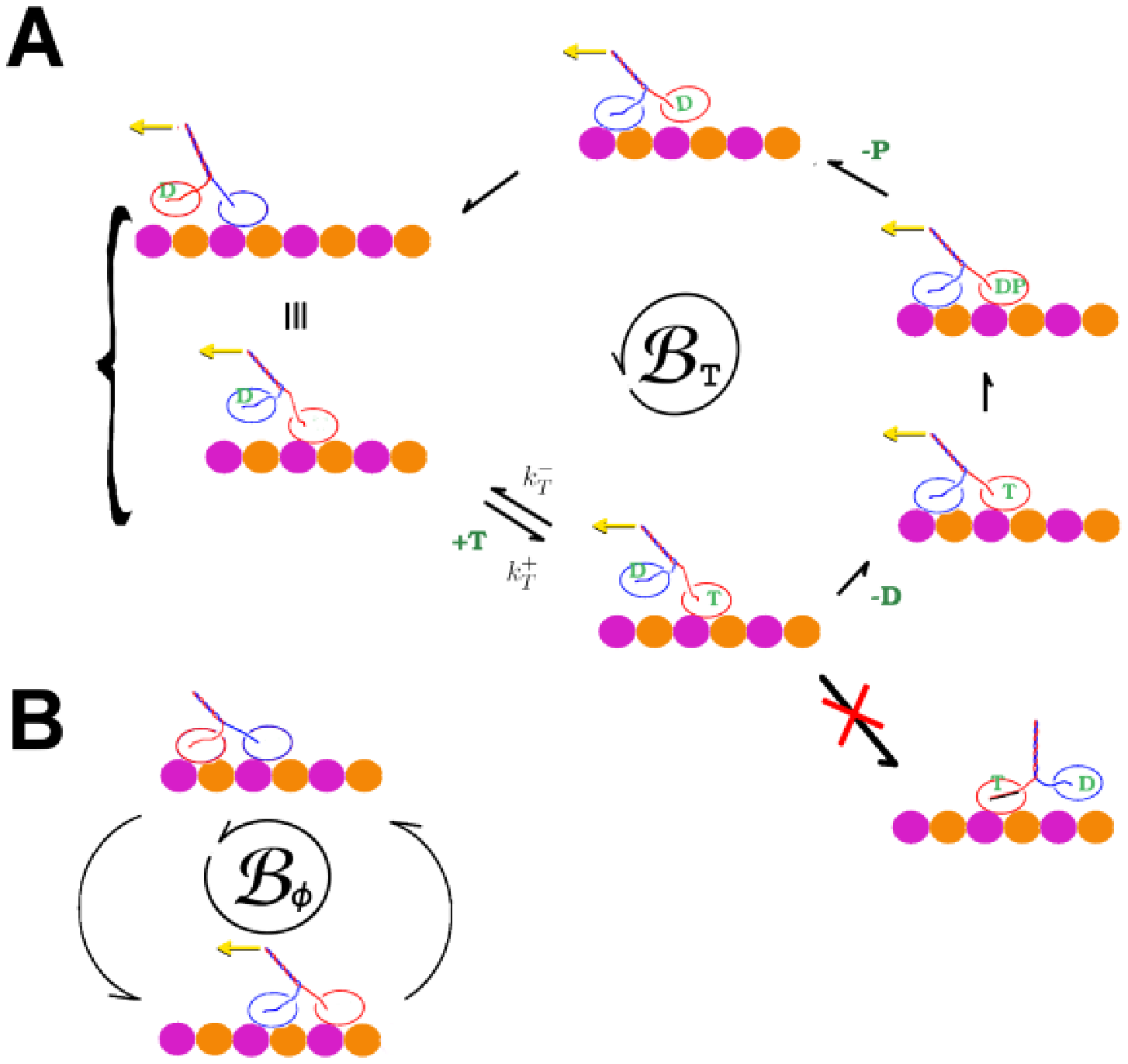}
   \caption{Two plausible cycles of kinesin backstep under tension. {\bf A}. ATP-induced backstepping cycle ($\mathcal{B}_T$). {\bf B.} Nucleotide-free backstepping cycle under tension ($\mathcal{B}_{\phi}$)}
  \end{center}
\end{figure}

\begin{figure}[ht]
  \begin{center}
   \includegraphics[width=5in]{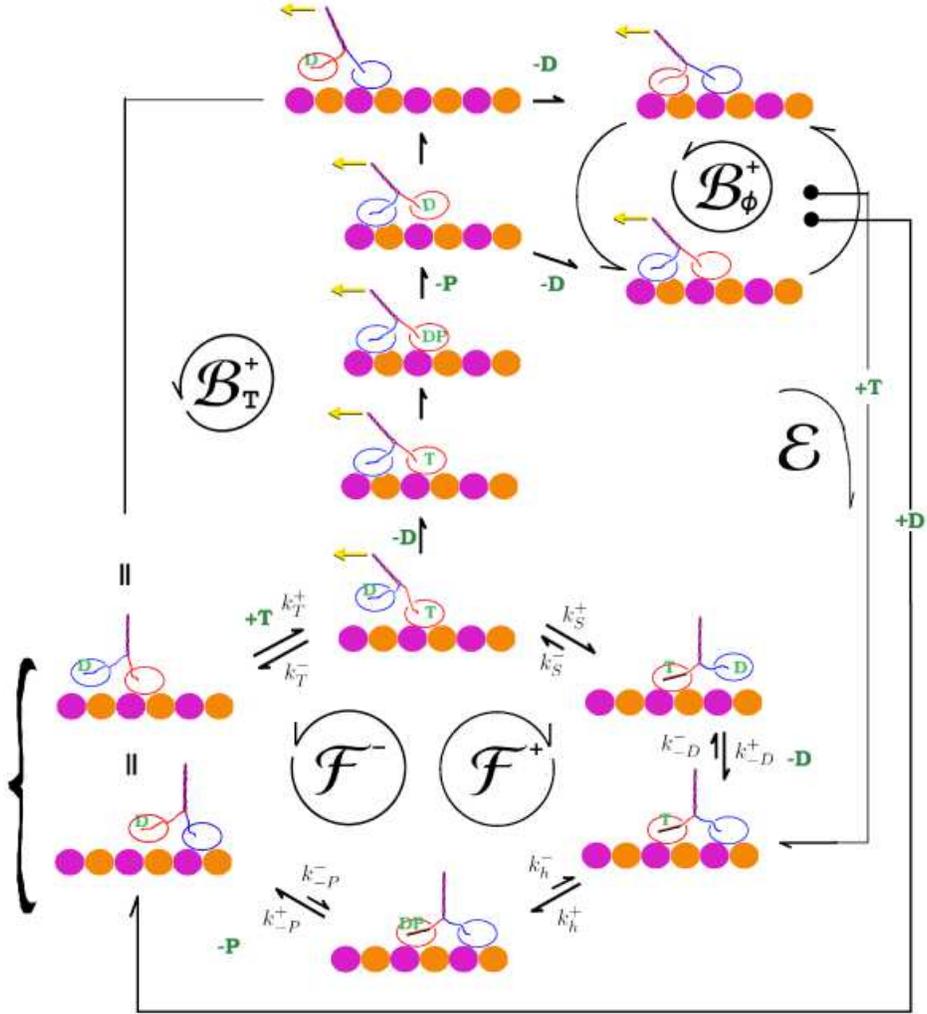}
   \caption{Expanded chemomechanical cycle of kinesins. The backstepping cycles $\mathcal{B}_T$ and $\mathcal{B}_{\phi}$ cycles are combined with the conventional single loop cycle $\mathcal{F}$. The motor escapes from $\mathcal{B}_{\phi}$, when ATP or ADP binds to the nucleotide free catalytic site of the trailing head. These two escape processes are indicated by $\mathcal{E}$.  }
  \end{center}
\end{figure}

\begin{figure}[ht]
  \begin{center}
   \includegraphics[width=4.5in]{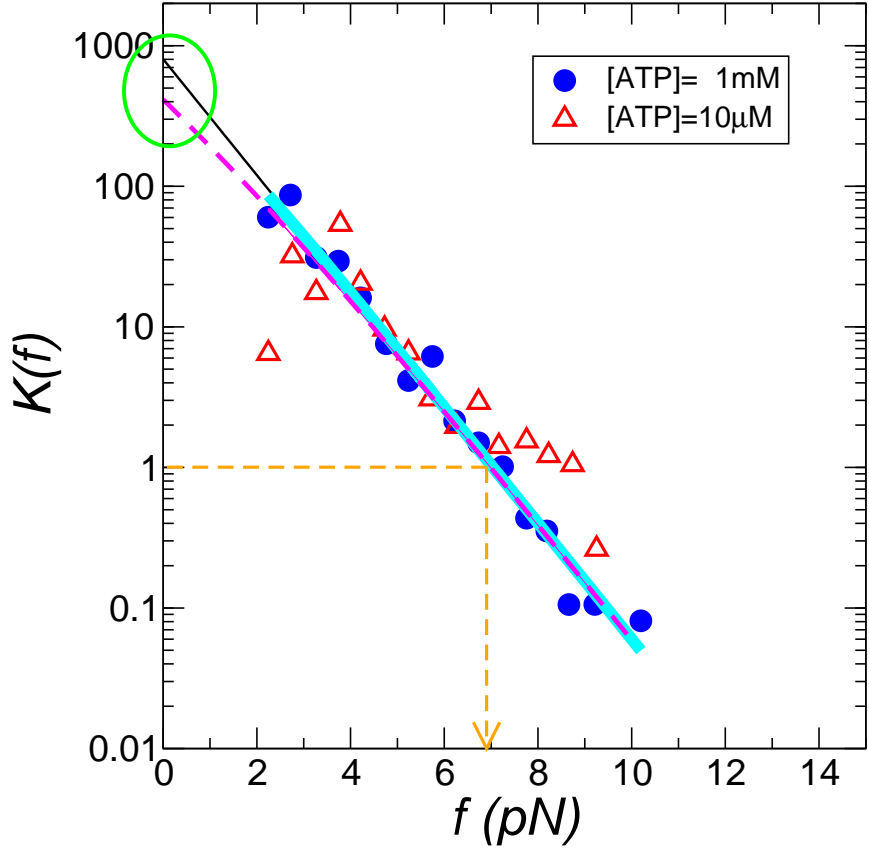}
   \caption{The force dependent ratio between forward and backward steps as obtained by Carter and Cross \cite{Cross05Nature}. The data were fitted to  
$802 e^{-0.95 f}/(1+K'e^{f\Delta/k_BT})$. 
The obtained fit parameters are  $(K',\Delta)=(0.79,-3.81)$ with a fit correlation coefficient 0.99 (thick solid line) for $[T]=1$ $mM$, and $(K',\Delta)=(0.94,-1.7)$ with a correlation coefficient 0.89 (dashed line) for $[T]=10$ $\mu M$. 
Contributions from the factor $K'e^{f\Delta/k_BT}$ in the denominator are practically negligible for large $f$ value, but we expect slightly downward shift in $K(f)$ value for $[T]=10$ $\mu M$ especially when $f\rightarrow 0$. 
The comparison between $K(f)=802\times e^{-0.95f}$ (thin solid line) and $K(f)=802\times e^{-0.95f}/(1+0.94e^{-1.7f/4.1})$ (dashed line) is made at $f\rightarrow 0$, enclosed in a circle.      
The stall force ($K(f)=1$) is $f\approx 7$ pN at both concentrations.}
  \end{center}
\end{figure}
\end{document}